\documentstyle[12pt]{article}

\newcommand{\eqref}[1]{eq.(\ref{#1})}

\newcommand{\beq}{\begin{equation}}
\newcommand{\eeq}{\end{equation}}
\newcommand{\bl}[1]{\makebox[#1em]{}}
\newcommand{\bm}[1]{\mbox{\boldmath$#1$}}
\newcommand{\pa}{\partial}
\newcommand{\ovl}[1]{\overline{#1}}
\newcommand{\ul}[1]{\underline{#1}}

\newcommand{\Qtil}{\hat{Q}}
\newcommand{\ftil}{\tilde{f}}
\newcommand{\gtil}{\tilde{g}}
\newcommand{\htil}{\tilde{h}}
\newcommand{\uhat}{\hat{u}}
\newcommand{\vhat}{\hat{v}}

\newcommand{\ii}{\mbox{i}}
\newcommand{\ee}{\mbox{e}}

\newenvironment{eqabc}%
{\setcounter{enumi}{\value{equation}}%
\addtocounter{enumi}{1}%
\setcounter{equation}{0}%
\begin{eqnarray}}%
{\end{eqnarray}\setcounter{equation}{\value{enumi}}}%

\begin{document}

\begin{titlepage}

\hfill
solv-int/9501005\\
\vspace{20pt}

\begin{center}
\begin{Large}
{\bf Bilinearization of a Generalized Derivative}\\
{\bf Nonlinear Schr\"odinger Equation}\\
\end{Large}

\vspace{35pt}

\noindent
Saburo Kakei,\footnote{e-mail: kakei@mmm.t.u-tokyo.ac.jp}
Narimasa Sasa and Junkichi Satsuma$^{\dag}$

\vspace{25pt}
\begin{small}
{\it Department of Applied Physics, Faculty of Engineering,
University of Tokyo,}\\
{\it Hongo 7-3-1, Bunkyo-ku, Tokyo 113, Japan}\\
$^{\dag}${\it Department of Mathematical Sciences, University of Tokyo,}\\
{\it Komaba 3-8-1, Meguro-ku, Tokyo 153, Japan}
\end{small}

\vspace{80pt}
{\bf Abstract}
\end{center}
\bl{2} A generalized derivative nonlinear Schr\"odinger equation,
\[
\ii q_t + q_{xx} + 2\ii \gamma |q|^2 q_x + 2\ii (\gamma-1)q^2 q^*_x
+ (\gamma-1)(\gamma-2)|q|^4 q = 0 ,
\]
is studied by means of Hirota's bilinear formalism. Soliton solutions
are constructed as quotients of Wronski-type determinants.
A relationship between the bilinear structure and gauge
transformation is also discussed.\\

\end{titlepage}

\section{Introduction}
The nonlinear Schr\"odinger (NLS) equation is one of the most generic
soliton equations. Various modifications of the equation have been
proposed. One of these attempts is to study the effects of
higher order perturbations.
Extended version of the NLS equation with higher order nonlinearity has
been proposed by various authors. It has been revealed that there
exist ``integrable'' cases even for extended equations.

Among them, there are two celebrated equations with derivative-type
nonlinearities, which are called the derivative nonlinear
Schr\"odinger (DNLS) equations.
One is the Chen-Lee-Liu (CLL) equation\cite{CLL,Nak},
\beq
\ii Q_T + Q_{XX} + 2\ii |Q|^2 Q_X = 0,
\label{eqn:CCL}
\eeq
and the other is the Kaup-Newell (KN) equation\cite{KN},
\beq
\ii Q_T + Q_{XX} + 2\ii (|Q|^2 Q)_X =0.
\eeq
It is known that the CLL and the KN equations may be
transformed into each other by a ``gauge transformation''\cite{WS}.
The method of gauge transformation can be applied to some generalized
case.  Kundu\cite{Ku} proposed the following equation:
\beq
\ii Q _T + Q _{XX} + \beta |Q|^2 Q
 + \ii (4\delta+2\alpha)|Q|^2 Q_X
 + \ii (4\delta+\alpha) Q^{2}Q^* _X
 + \delta(4\delta+\alpha) |Q|^4 Q =0,
\label{eqn:GDNLS0}
\eeq
where $Q^*$ denotes the complex conjugate of $Q$.
If we set
\beq
Q=\Qtil \exp\left( -2\ii\delta\int^{X}|\Qtil|^2 \mbox{d}X \right),
\label{eqn:GT}
\eeq
then \eqref{eqn:GDNLS0} is gauge-equivalent to
\beq
\ii \Qtil_T + \Qtil_{XX} + \beta|\Qtil|^2 \Qtil
 + \ii\alpha\left( |\Qtil|^2 \Qtil \right)_X = 0,
\eeq
which is a hybrid of the NLS equation and the DNLS
equation of KN-type.
In case of $\alpha > 0$,
\eqref{eqn:GDNLS0} is transformed into a normalized form;
\beq
\ii q_t + q_{xx} + 2\ii \gamma |q|^2 q_x + 2\ii (\gamma-1)q^2 q^*_x
+ (\gamma-1)(\gamma-2)|q|^4 q = 0 ,
\label{eqn:GDNLS}
\eeq
by means of the change of variables,
\[
Q(X,T) \;=\; \sqrt{\frac{2}{\alpha}}q(x,t)
\exp\left(\frac{\ii\beta}{\alpha}x
+ \frac{\ii\beta^2}{\alpha^2}t
\right),
\] \[ X \;=\; x+\frac{2\beta}{\alpha}t, \bl{2} T \;=\; t, \bl{2}
\gamma \;=\; \frac{4\delta}{\alpha}+2 .
\]
Clarkson and Cosgrove applied the Painlev\'e test to this type of
equations and check the integrability\cite{Cla}.

The aim of this paper is to construct multi-soliton solutions for
\eqref{eqn:GDNLS}. Soliton solutions for the KN equation ($\gamma =2$)
are already known\cite{KN},
and solutions for \eqref{eqn:GDNLS} are constructed by
the gauge transformation (\ref{eqn:GT}) in principle. However, to obtain
the explicit form of the solutions, one must integrate
\eqref{eqn:GT} in practice. The integration becomes very complicated
in multi-soliton case.
In this paper, we study \eqref{eqn:GDNLS} by means of Hirota's
bilinear formalism, and construct soliton solutions through the
Wronskian technique\cite{HOS}.
One can get clear insights for the algebraic structure
through the bilinear method.

\section{Bilinearization}
In ref.2, Nakamura and Chen bilinearized the CLL equation
($\gamma=1$) and constructed soliton solutions.
Hirota found that the CLL equation and the KN equation ($\gamma=2$)
can be transformed into the same bilinear equations\cite{Hirota}.
Here we generalize their results and show that \eqref{eqn:GDNLS} shares
the same bilinear structure for all $\gamma$.

For the moment, we will forget the complex structure of
\eqref{eqn:GDNLS}, and start with the coupled equations;
\begin{eqabc}
u_{x_2} - u_{x_1 x_1} + 2\gamma uvu_{x_1} + 2(\gamma-1)u^2 v_{x_1}
+ (\gamma-1)(\gamma-2)u^3 v^2 & = & 0, \\
-v_{x_2} - v_{x_1 x_1} -2\gamma uv v_{x_1} - 2(\gamma-1)v^2 u_{x_1}
+ (\gamma-1)(\gamma-2)u^2 v^3 & = & 0.
\end{eqabc}%
Note that in case of $v=\pm u^*$, eqs.(7) are reduced to
\eqref{eqn:GDNLS} by setting $q=u$, $x_1 = \pm \ii x$ and
$x_2 = -\ii t$. Using the dependent variable transfomations,
\beq
u \;=\; \frac{f^{\gamma-1}g}{\ftil^\gamma}, \bl{2}
v \;=\; \frac{\ftil^{\gamma-1}\gtil}{f^\gamma},
\label{eqn:DVT}
\eeq
eqs.(7) may be decoupled into the following bilinear equations;
\begin{eqabc}
(D_{x_1}^2 + D_{x_2}) f \cdot \ftil & = & 0, \label{eqn:bilinear1}\\
(D_{x_1}^2 - D_{x_2}) f \cdot \gtil & = & 0, \\
(D_{x_1}^2 + D_{x_2}) \ftil \cdot g & = & 0, \\
D_{x_1}^2 f \cdot f + 2gh & = & 0, \\
D_{x_1}^2 \ftil \cdot \ftil + 2\gtil\htil & = & 0, \\
D_{x_1} f \cdot \ftil - g\gtil & = & 0, \label{eqn:bilinear6}\\
D_{x_1} f \cdot \gtil - \ftil h & = & 0, \\
D_{x_1} \ftil \cdot g + f \htil & = & 0,
\label{eqn:bilinear}
\end{eqabc}%
where we have introduced auxiliary independent variables $h$ and
$\htil$.
In eqs.(9) we use the $D$-operator of Hirota\cite{HS},
defined by
\[
D_x^m D_t^n F(x,y)\cdot G(x,t) \;=\;
\left. ( \pa_x-\pa_{x'})^m ( \pa_t-\pa_{t'})^n
F(x,y)G(x',t') \right|_{x'=x,t'=t}.
\]
On decoupling eqs.(7) into eqs.(9), the following properties of
the $D$-operator, valid for any functions $F(x)$, $G(x)$ and $H(x)$,
are useful;
\begin{eqabc}
\frac{\pa}{\pa x}\left(\frac{F^{\gamma-1}G}{H^\gamma}\right) & = &
\frac{F^{\gamma-1}}{H^{\gamma+1}}D_x G\cdot H \;+\;
(\gamma-1)\frac{F^{\gamma-2}G}{H^{\gamma+1}}D_x F\cdot H ,
\label{eqn:Dop1}\\
\frac{\pa^2}{\pa x^2}\left(\frac{F^{\gamma-1}G}{H^\gamma}\right) & = &
\frac{F^{\gamma-1}}{H^{\gamma+1}}D_x^2 G\cdot H \;+\;
(\gamma-1)\frac{F^{\gamma-2}G}{H^{\gamma+1}}D_x^2 F\cdot H \nonumber \\
\lefteqn{ \;-\; \gamma\frac{F^{\gamma-1}G}{H^{\gamma+2}}D_x^2 H\cdot H
   \;+\; 2(\gamma-1)\frac{F^{\gamma-2}}{H^{\gamma+2}}(D_x F\cdot H)
   (D_x G\cdot H)}
   \bl{3} \nonumber\\
\lefteqn{ \;+\; (\gamma-1)(\gamma-2)\frac{F^{\gamma-3}G}{H^{\gamma+2}}
(D_x F\cdot H)^2,} \bl{3}
\label{eqn:Dop2}\\
D_x^2 F\cdot G & = & \frac{G}{2F}D_x^2 F\cdot F \;+\;
\frac{F}{2G}D_x^2 G\cdot G \;+\; \frac{1}{FG}(D_x F\cdot G)^2 .
\end{eqabc}%
Relations (10a, b) are obtained
from an expansion of the following identity;
\[
\exp\left(\epsilon\frac{\pa}{\pa x}\right)
\frac{F^{\gamma-1}G}{H^\gamma} \;=\;
\frac{\left(\exp\left(\epsilon D_x\right)F\cdot H\right)^{\gamma-1}
\left(\exp\left(\epsilon D_x\right)G\cdot H\right)}{
\left(\mbox{cosh}\left(\epsilon D_x\right)G\cdot H\right)^{\gamma}},
\]
and (10c) is from
\[
\left( \exp(\epsilon D_x)F\cdot F \right)
\left( \exp(\epsilon D_x)G\cdot G \right)
\;=\; \left( \exp(\epsilon D_x)F\cdot G \right)
\left( \exp(\epsilon D_x)G\cdot F \right) .
\]
The latter identity is a special case of eq.(VII) of ref.9
($\delta=0$, $a=b$, $c=d$).
Additional formulas listed in the appendix of ref.9 are also
useful.

We note that eqs.(9) are overdetermined. However, once we have a set of
solutions for eqs.(9), we can construct a solution of eqs.(7) through the
transformation (8)

\section{Construction of Solutions}
It is known that solutions of bilinear equations are
expressed as quotients of Wronski-type determinants\cite{HOS}.
We will show that solutions of the bilinear equations (9) can also be
constructed by the Wronskian technique.

In the following, we use abbreviation,
\beq
\ovl{n} \;=\; \left( \frac{\pa}{\pa x_1^{(1)}} \right)^{n}
\left( \begin{array}{c}
\varphi_1(x^{(1)}) \\ \varphi_2(x^{(1)}) \\ \vdots \\ \varphi_{2N}(x^{(1)})
\end{array} \right), \bl{2}
\ul{n} \;=\; \left( \frac{\pa}{\pa x_1^{(2)}} \right)^{n}
\left( \begin{array}{c}
\psi_1(x^{(2)}) \\ \psi_2(x^{(2)}) \\ \vdots \\ \psi_{2N}(x^{(2)})
\end{array} \right).
\eeq
Define the {\it $\tau$-function},
\beq
\tau_{m,n;i} \;=\; \left| \ovl{m},\ovl{m+1},\cdots,\ovl{m+N-1-i};
\ul{n},\ul{n+1},\cdots,\ul{n+N-1+i} \right| ,
\label{tau}
\eeq
where the elements of the determinant satisfy the linear differential
equations ({\it dispersion relations});
\beq
\frac{\pa \varphi_j}{\pa x_n^{(i)}} \;=\;
\delta_{i1}\frac{\pa^{n} \varphi_j}{\pa x_1^{(1)n}}, \bl{2}
\frac{\pa \psi_j}{\pa x_n^{(i)}} \;=\;
\delta_{i2}\frac{\pa^{n} \psi_j}{\pa x_1^{(2)n}}.
\label{eqn:DispRel}
\eeq
This type of determinant is called ``{\it
double-Wronskian}''\cite{HOS}.

Hereafter we assume this $\tau$-function has the
$A^{(1)}_1$-symmetry\cite{JM};
\beq
\tau_{m,n;i} = \tau_{m+1,n+1;i}.
\label{eqn:reduction}
\eeq
Solutions for the overdetermined equations (9) can be constructed owing to
this symmetry.
Under this assumption, the $\tau$-function depends only on the
difference of $x_j^{(1)}$ and $x_j^{(2)}$ for all $j$. Set
\beq
\begin{array}{rclcrcl}
f & = & \tau_{0,0;0}, & \bl{1} & \ftil & = & \tau_{1,0;0}, \\
g & = & \tau_{0,0;-1}, & \bl{2} & \gtil & = & \tau_{1,0;1}, \\
h & = & \tau_{0,0;1}, & \bl{2} & \htil & = & \tau_{1,0;-1},
\end{array}
\label{eqn:solutions}
\eeq
and
\beq
x_1 \;=\; x_1^{(1)}-x_1^{(2)}, \bl{2}
x_2 \;=\; x_2^{(1)}-x_2^{(2)}.
\label{eqn:XandT}
\eeq
Then one can show that $f$, $g$, $h$, $\ftil$, $\gtil$ and $\htil$
above satisfy eqs.(9). Here we only prove the first of
eqs.(9) by using the Laplace expansion.
Let us consider the following identity for $4N \times 4N$ determinant,
\beq
\left| \begin{tabular}{c|c|c|c|c|c}
$\ovl{0}$ & $\ovl{1} ,\cdots, \ovl{N-2}$ & $\emptyset$ &
$\ovl{N-1},\ovl{N},\ovl{N+1}$ & $\ul{0} ,\cdots, \ul{N-1}$ & $\emptyset$ \\
\hline
$\emptyset$ & $\emptyset$ & $\ovl{1} ,\cdots, \ovl{N-2}$ &
$\ovl{N-1},\ovl{N},\ovl{N+1}$ & $\emptyset$ & $\ul{0} ,\cdots, \ul{N-1}$
\end{tabular} \right| \;=\; 0.
\label{eqn:4N*4N}
\eeq
Applying the Laplace expansion into $2N \times 2N$ minors to the
left-hand side of \eqref{eqn:4N*4N}, one finds
\begin{eqnarray}
\lefteqn{|\ovl{0},\cdots,\ovl{N-2},\ovl{N-1};\ul{0},\cdots,\ul{N-1}|}
\bl{1} \nonumber \\
 & \times &
|\ovl{1},\cdots,\ovl{N-2},\ovl{N},\ovl{N+1};\ul{0},\cdots,\ul{N-1}|
\nonumber \\
\lefteqn{-\;|\ovl{0},\cdots,\ovl{N-2},\ovl{N};\ul{0},\cdots,\ul{N-1}|}
\bl{1} \nonumber \\
 & \times &
|\ovl{1},\cdots,\ovl{N-2},\ovl{N-1},\ovl{N+1};\ul{0},\cdots,\ul{N-1}|
\nonumber \\
\lefteqn{+\;|\ovl{0},\cdots,\ovl{N-2},\ovl{N+1};\ul{0},\cdots,\ul{N-1}|}
\bl{1} \nonumber \\
 & \times &
|\ovl{1},\cdots,\ovl{N-2},\ovl{N-1},\ovl{N};\ul{0},\cdots,\ul{N-1}|
\;=\;0 .
\label{eqn:Plucker1}
\end{eqnarray}
{}From \eqref{eqn:Plucker1}, we have
\beq
( D_{x_1^{(1)}}^2 + D_{x_2^{(1)}} )
\tau_{0,0;0} \cdot \tau_{1,0;0} \;=\; 0.
\label{eqn:bilin1}
\eeq
On the other hand, if we apply the Laplace expansion to
\beq
\left| \begin{tabular}{c|c|c|c|c|c}
$\ovl{1} ,\cdots, \ovl{N}$ & $\emptyset$ & $\ul{0}$ &
$\ul{1} ,\cdots, \ul{N-2}$ & $\emptyset$ & $\ul{N-1},\ul{N},\ul{N+1}$ \\
\hline
$\emptyset$ & $\ovl{1} ,\cdots, \ovl{N}$ & $\emptyset$ &
$\emptyset$ & $\ul{1} ,\cdots, \ul{N-2}$ & $\ul{N-1},\ul{N},\ul{N+1}$
\end{tabular} \right| \;=\; 0,
\eeq
we get
\beq
( D_{x_1^{(2)}}^2 + D_{x_2^{(2)}} )
\tau_{1,0;0} \cdot \tau_{1,1;0} \;=\; 0.
\label{eqn:bilin2}
\eeq
Under the conditions (\ref{eqn:reduction})--(\ref{eqn:XandT}), both
eqs.(\ref{eqn:bilin1}) and (\ref{eqn:bilin2}) reduce to
\eqref{eqn:bilinear1}.
The remaining equations in eqs.(9) can be proved in the same way.

In order to construct soliton solutions of \eqref{eqn:GDNLS},
one may choose $\varphi_j (x^{(1)})$ and $\psi_j (x^{(2)})$ in
eqs.(13) as
\begin{eqabc}
\varphi_j(x^{(1)}) & = & a_j \exp(p_j x_1^{(1)} + p_j^2 x_2^{(1)}),\\
\psi_j(x^{(2)}) & = & b_j \exp(p_j x_1^{(2)} + p_j^2 x_2^{(2)}).
\end{eqabc}%
If we impose the conditions,
\beq
p_{N+j} \;=\; p_j^*, \bl{2} |p_j|^2 \;=\; 1, \bl{2}
\frac{b_{N+j}}{a_{N+j}} \;=\;
\left( \frac{a_j}{b_j} p_j \right)^* , \bl{2}j=1,2,\cdots,N,
\label{eqn:TheConditions}
\eeq
on $\varphi_j (x^{(1)})$ and $\psi_j (x^{(2)})$ above,
and assume $x_1^{(1)}$, $x_1^{(2)}$, $x_2^{(1)}$ and $x_2^{(2)}$ are
pure imaginary, then
the functions $f$, $g$, $\ftil$ and $\gtil$ defined by
eqs.(\ref{eqn:solutions}) satisfy the relations;
\beq
\left( \frac{g}{f} \right)^* \;=\; -\frac{\gtil}{\ftil}, \bl{2}
\left( \frac{f}{\ftil} \right)^* \;=\; \frac{\ftil}{f}.
\label{eqn:reality}
\eeq
(The reduction condition (\ref{eqn:reduction}) is satisfied
under the assumptions (22), (\ref{eqn:TheConditions}).)
Hence we have $v=-u^*$.

To prove the reality condition (\ref{eqn:reality}), we start with the
linear equation;
\beq
\left( \ovl{0},\ovl{1},\cdots,\ovl{N-1};
\ul{0},\ul{1},\cdots,\ul{N-1} \right) \bm{w}_1 \;=\;
\left( \ovl{N} \right) ,
\label{eqn:linEq1}
\eeq
for
\[
\bm{w}_1 \;=\; \left(
\begin{array}{c}
w_1^{(1)} \\ w_1^{(2)} \\ \vdots \\ w_1^{(2N)}
\end{array} \right).
\]
We find that $w_1^{(2N)} = (-)^{N-1}g/f$ by solving \eqref{eqn:linEq1}.
Next, by solving
\beq
\left( \ovl{1},\ovl{2},\cdots,\ovl{N};
\ul{0},\ul{1},\cdots,\ul{N-1} \right) \bm{w}_2 \;=\;
\left( \ul{N} \right) ,
\label{eqn:linEq2}
\eeq
for
\[
\bm{w}_2 \;=\; \left(
\begin{array}{c}
w_2^{(1)} \\ w_2^{(2)} \\ \vdots \\ w_2^{(2N)}
\end{array} \right),
\]
we get $w_2^{(N)} = (-)^{N}\gtil /\ftil $.
Under the conditions (\ref{eqn:TheConditions}),
\eqref{eqn:linEq1} is transformed into the complex conjugate of
\eqref{eqn:linEq2} by suitable rearrangement of rows and columns.
By using this fact, we
obtain the first one of eqs.(\ref{eqn:reality}).
The second is similarly proved, but in this case we use
the symmetry (\ref{eqn:reduction}).

As an example of the solutions, we give the explicit form of the
$1$-soliton solution,
\beq
q(x,t) \;=\; \frac{p-p^*}{c\ee^{\phi}+\frac{1}{c^* p}\ee^{-\phi^*}}
\left( \frac{
c\ee^{\phi}+\frac{1}{c^* p}\ee^{-\phi^*}
}{
cp^*\ee^{\phi}+\frac{1}{c^*}\ee^{-\phi^*}} \right)^{\gamma},
\label{eqn:1soliton}
\eeq
where $c$ and $p$ are complex constants. The parameter $p$ is chosen to
satisfy $|p|^2 =1$. The phase $\phi(x,t)$ is defined by
\[
\phi(x,t) \;=\; \ii p x + \ii p^2 t + \phi^{(0)},
\]
with $\phi^{(0)}$ an arbitrary constant.
If $\gamma$ is an integer, \eqref{eqn:1soliton} becomes rational w.r.t.
exponential functions, as is always the case of soliton solutions.
But in general case,
\eqref{eqn:1soliton} is {\it not} rational. This type of soliton
solution has not been obtained so far as the authors know.

\section{Relation with the Gauge Transformation}
In this section, we discuss the relationship between the gauge
transformation method\cite{WS,Ku} and our theory.

We set $\uhat$ and $\vhat$ as a solution of eqs.(7) with
$\gamma=2$,
\begin{eqabc}
\uhat_{x_2} - \uhat_{x_1 x_1} + 4\uhat\vhat\uhat_{x_1}
 + 2\uhat^2 \vhat_{x_1} & = & 0, \\
-\vhat_{x_2} - \vhat_{x_1 x_1} -4\uhat\vhat \vhat_{x_1}
 - 2\vhat^2 \uhat_{x_1} & = & 0.
\end{eqabc}%
As is shown in the preceding section, the solution of the equations above
is represented as eqs.(\ref{eqn:DVT}) with $\gamma =2$;
\beq
\uhat \;=\; \frac{f g}{\ftil^2}, \bl{2}
\vhat \;=\; \frac{\ftil \gtil}{f^2}.
\label{eqn:DVT2}
\eeq
In this case, the gauge transformation between eqs.(28) (eqs.(7) with
$\gamma =2$) and eqs.(7) (arbitrary $\gamma$) is written as
\begin{eqabc}
u & = & \uhat \exp\left( -(\gamma-2)\int^{x_1}
\uhat\vhat \mbox{d}x_1 \right),\\
v & = & \vhat \exp\left( (\gamma-2)\int^{x_1}
\uhat\vhat \mbox{d}x_1 \right).
\end{eqabc}%
If we substitute $u$, $v$ of eqs.(\ref{eqn:DVT}) and $\uhat$, $\vhat$ of
eqs.(\ref{eqn:DVT2}), eqs.(30) are transformed into
\beq
\int^{x_1} \frac{g\gtil}{f\ftil} \mbox{d}x_1 \;=\; \log\frac{\ftil}{f}.
\label{eqn:GTandBL}
\eeq
Differentiating both side of \eqref{eqn:GTandBL}, one finds
\beq
\frac{g\gtil}{f\ftil} \;=\; \frac{D_{x_1}f\cdot\ftil}{f\ftil}.
\label{eqn:GTandBL2}
\eeq
Equation (\ref{eqn:GTandBL2}) is nothing but \eqref{eqn:bilinear6}.
Hence we conclude that the gauge transformation (30) is a direct
consequence of the bilinear structure of the equations.
We remark that the bilinear variables, $f$, $g$, $\ftil$ and $\gtil$,
clarify the structure of the gauge transformation.

\section*{Acknowledgement}
We are grateful to Dr. Kenji Kajiwara, Mr. Kai Matsui, Dr. Yasuhiro
Ohta and Dr. Tetsu Yajima for fruitful discussions and helpful
comments.

\end{document}